\documentclass[aps,amsmath,pre,showpacs]{revtex4}

\usepackage{graphicx}
\usepackage{amsmath}

\def\(({\left(}
\def\)){\right)}
\def\[[{\left[}
\def\]]{\right]}

\def \form#1 {eq. (\ref{#1}) }
\def \parziale#1#2  {{\partial {#1} \over \partial {#2}}}

\def \cN{{\cal N}}

\def\la{\langle}
\def\ra{\rangle}

\def \ba#1 {\overline{#1}}

\def  \s {\sigma}

\begin{document}
\title{Near optimal configurations in mean field disordered systems}

\author{A. Pagnani}
\affiliation{Laboratoire de Physique Th\'eorique et Mod\`eles Statistiques,
b\^at. 100, Universit\'e Paris-Sud, F--91405 Orsay, France.}
\author{G. Parisi}
\affiliation{Dipartimento di Fisica, SMC, INFM, and INFN, Universit\`a
di Roma 1 {\em La Sapienza, P.le A. Moro, 2 -- 00185 Roma, Italy.}} 
\author{M. Rati\'eville}
\affiliation{Dipartimento di Fisica, SMC, INFM, Universit\`a di Roma 1
{\em La Sapienza, P.le A. Moro, 2 -- 00185 Roma, Italy}} 
\affiliation{Laboratoire de Physique Th\'eorique et Mod\`eles Statistiques
Universit\'e Paris Sud -- {\em 91405 Orsay, France.}}

\begin{abstract}

We present a general technique to compute how the energy of a
configuration varies as a function of its overlap with the ground
state in the case of optimization problems. Our approach is based on a
generalization of the cavity method to a system interacting with its
ground state.  With this technique we study the random matching
problem as well as the mean field diluted spin glass. As a byproduct
of this approach we calculate the de~Almeida-Thouless transition
line of the spin glass on a fixed connectivity random graph.

\end{abstract}


\maketitle

\section{Introduction}

The study of the ground state properties of disordered systems reveals
deep connections with the field of random combinatorial optimization:
in fact combinatorial optimization problems can be stated in terms of
zero-temperature statistical properties of some disordered system
Hamiltonian. A closely related issue is the {\em computational
complexity} of a given problem that can be stated in terms of the
typical amount of time ({\em e.g.} cpu-cycles) that is needed to solve
the problem as a function of its size \cite{mertens}. Easy/Hard
thresholds have been observed in different combinatorial optimization
problems \cite{ksat,coloring,vertex,number}, and a currently highly
debated issue is the interplay between the onset of the phase
transition and the slowing down of local search algorithms
\cite{monasson}. A better understanding of the organization of the
lowest energy configurations is then a promising research program not
only to relate the {\em computational complexity} of a given problem
to its statistical mechanics counterpart, but also as a guide for the
implementation of more efficient algorithms.

Recently the problem of diluted mean field spin glasses received a
renewed interest by means of the {\em cavity approach}
\cite{MP01,MP02}. Interestingly this method turned out to be a
powerful tool also to deal with combinatorial optimization
problems. In this paper we study the problem of the organization of the
lowest energy configurations within this framework. Our aim is to explore the
way the energy of a configuration varies as a function of its overlap
with the ground state. The naive procedure consisting in computing the
ground state, then picking at random configurations and look at their
distance to the ground state is not computationally practicable:
configurations are far too numerous for this sampling to be efficient,
let alone that one has to do it for numerous instances of the coupling
constants in order to get relevant averaged quantities. We need to
generate configurations whose energy density is somewhat under
control. The technique we use consists in perturbing the Hamiltonian
to increase the energy of the ground state, then compute the new
ground state. The perturbation is chosen proportional to a small
parameter $\varepsilon$ which we can tune to make the new ground state more or less distant from the original one: hereafter we will
refer to this method as $\varepsilon$-coupling. This is actually not a
new idea \cite{palassini,marinari-parisi,hartmann,franz-parisi, RNA},
however to our knowledge so far this technique has been implemented
only in numerical simulations. Here we show that an analytic solution
is possible in the case of the simple random matching problem as well
as the case of spin glasses on random graphs of fixed connectivity.

In a recent paper Aldous and Percus \cite{aldous_percus} used similar
techniques to study the matching problem (both in the Euclidean and
mean field versions) and the traveling salesman problem, conjecturing
that it is possible to classify optimization problems into different
universality classes according to the dependence of the ground state
solution on small perturbations. We will compare our results to
those presented in \cite{aldous_percus}, and we will explain how the
Replica Symmetry Breaking (RSB) transition can be detected within the
cavity scheme. Let us point out that we will restrict our analysis to
the level of the replica symmetric (RS) approximation although in
principle there is no problem to extend the same analysis to higher
level of replica symmetry breaking; we will discuss this possibility
and the potential interest of this generalization in the context of
other combinatorial optimization problems ({\em e.g.} SAT \cite{ksat}
and coloring \cite{coloring}).

The rest of the paper has the following structure. In
Sec.~\ref{sec:matching} we study the random simple matching
problem: after a description of the model we show how the cavity
approach works in general for this model and we introduce a
generalization in order to deal with $\varepsilon$-coupled systems.
In Sec.~\ref{sec:gaussian_SG} we explain how the method works in the
case of the diluted spin glass where a RSB
transition is known to exist. In Sec.~\ref{sec:concl} some
conclusions and perspectives are presented.

\section{The random simple matching problem}
\label{sec:matching}
\subsection{The Model}
Given an unoriented graph $G=(V,E)$, where $V$ are the vertices and
$E$ are the edges, a {\em matching} $M$ is a set of edges having the
property that no two edges in $M$ have an end in common. We say that a
vertex $v\in V$ is {\em matched} if there is an edge incident to $v$
in the matching. Otherwise the vertex is {\em unmatched}. A matching
is called {\em perfect} if every vertex of $G$ is matched.

In the following we stick to the case where $G$ is {\em complete},
i.e. there is an edge between any two vertices, and when we speak of a
matching we mean a perfect matching.

We introduce a function $l$ defined on the set of edges $E$ which
associates a real number $l(e)$ to each edge $e\in E$. The quantity
$l(e)$ can be thought of as a distance, or -- depending on the taste
-- as a weight, a cost, etc. We define the total length of a matching
$M$ as $L_M=\sum_{e\in M} l(e)$. The matching problem consists in
finding the cheapest matching, i.e. which minimizes $L_M$.

A classic example is the random Euclidean matching problem: the
vertices are identified with $N$ points drawn at random with the flat
measure in the unit hypercube of a $d$-dimension Euclidean space. The
cost $l(e)$ of an edge $e=\{v,w\}$ is the usual Euclidean distance
between the two vertices $v$ and $w$.

A mean field approximation of the Euclidean problem has been widely
investigated
\cite{mezpar2,mezpar,mezpar3,houdayer,brunetti,ratieville2}: the
weights of the edges are i.i.d. random variables whose common
probability distribution $\rho$ is defined on the interval
$[0,+\infty[$. The function $\rho(l)$ is assumed to behave like an
integer power law for small $l$:
\begin{equation}
\label{powerlaw}
\rho(l)  \sim \frac{l^r}{r!}.  
\end{equation} 
In the thermodynamic limit $N\rightarrow \infty$ limit, the mean
distance of two nearest neighbors goes to 0 like $N^{-\delta}$, where
$\delta=1/(r+1)$. Intuitively a minimum matching will only include
edges of this order of magnitude, so that the only relevant feature of
$\rho$ is its behavior around 0, i.e. the $r$ exponent.
%Therefore from the thermodynamic
%point of view the model is characterized by how $\rho$ tends to 0 for
%small distances ({\em i.e.} by $r$).
The one-edge and two-edge length distributions in this model match the
ones of the Euclidean random matching problem in dimension $d=r+1$ for
short distances.

%Indeed in the Euclidean case one can do the small
%following computation: let $l_{12}$ be the distance between two points
%and $\rho_E$ its probability law. For short $l$, one has
%\begin{equation}P
%(l_{12}<l)=\int_0^l  \rho_E(l)dl=B_d l^r, 
%\end{equation} 
%where $B_d$ is the volume of the unit ball in $d$ dimensions. So
%$\rho_E(l)=B_d l^{r-1}$. The mean field random matching problem of
%parameter $r$ is the best mean field approximation to the Euclidean
%matching in dimension $d=r+1$. This model has been extensively studied
%\cite{mezpar2,mezpar,brunetti, ratieville2} in the last years. A
%natural question is: how good an approximation is it? It will not be
%our concern here: we will consider the mean field random matching
%problem as an independent problem in its own right. Let us just
%mention that quantitative comparisons between the mean field and the
%Euclidean random matching problems have been carried out
%\cite{houdayer}, as well as an analytic correction to the mean field
%model to include the correlations between three edges
%\cite{euclidean}.

Hereafter we will concentrate on the case $r=0$. We will use the
following conventions: there are $N$ vertices; the distances
$l_{ij}=l_{ji}$ between two vertices are i.i.d. random variables
distributed following the flat distribution on $[0,N]$ (this
corresponds to a rescaling of a factor $N$ with respect to
Eq.(\ref{powerlaw})); as a shorthand notation, we indicate the set of
these coupling constants as $\ell$.  The length of the minimum
matching is an extensive quantity, and the energy (Hamiltonian) of a
matching can be defined equal to its length. A matching $M$ can be
univoquely represented by a contact matrix $(n_{ij})$ such that:
\begin{itemize}
\item  $n_{ij}\in\{0,1\}$ and $n_{ij}=n_{ji}$.
\item $n_{ii}=0$: no self linkage is allowed.
\item $\forall i,\, \sum_{j=1}^{N} n_{ij}=1$: each site cannot be
linked more than once. 
\end{itemize} 
Obviously $n_{ij}=0$ if the edge $\{i,j\}$ is not in $M$, and
$n_{ij}=1$ if the edge $\{i,j\}$ is in $M$. The entry $n_{ij}$ is
called the {\em occupation number} of the edge $\{i,j\}$. The
Hamiltonian of the matching then reads:
\begin{equation}
L_{\ell}=\sum_{1\leq i<j\leq N} n_{ij} l_{ij},
\end{equation}
where $\ell$ plays the role of the quenched disorder, and the $n_{ij}$
are the dynamic local variables.

\subsection{The cavity equations}

The cavity equations at finite temperature for the matching problem
have been derived in \cite{mezpar} (a comprehensive introduction to
this subject is found in \cite{houches}). Let us briefly reproduce the
basic steps following \cite{mez}. The partition function for the
matching problem is
\begin{equation}
\label{defZ}
Z = \sum_{\{n_{ij}\}} \exp \left( -\beta N \sum_{1\leq
i<j\leq N} n_{ij} l_{ij}\right)\,\,,
\end{equation}
where the scaling factor $N$ makes the free energy $F = -\log(Z)/\beta
$ an extensive quantity. Following polymers theory \cite{gennes}, one
first introduces a more tractable representation of the partition
function, which consists in mapping the matching problem onto a system
of interacting spins. On each vertex $i$ one puts a $p$-dimensional
vector spin ${\bf S}_i=(S_i^1,\ldots S_i^p)$ normalized by ${\bf
  S}_i^2=p$. Let $d\mu$ be the integration measure on the
corresponding sphere.  If we define the coupling constants
$T_{ij}=\exp \left( -\beta N l_{ij} \right)$, one can check that the
partition function (\ref{defZ}) can be written as:
\begin{equation}
\label{defz}
Z=\lim_{p \rightarrow 0} \int \left[ \prod_{i=1}^N S_i^1 d\mu({\bf
S}_i)\right] \exp \left( \sum_{1\leq i<j\leq N} T_{ij} {\bf
S}_i.{\bf S}_j \right)\,\,.
\end{equation}
Expanding the exponential into a power series and applying the following
property:
\begin{equation}
\label{loi}
\mathop{\lim}_{p \rightarrow 0} \int d\mu({\bf S}_i) S_i^{\alpha_1}
S_i^{\alpha_ 2} \cdots S_i^{\alpha_q}= 
\delta_{q,2}\,\delta_{\alpha_1,\alpha_2}
\end{equation}
one can easily recover (\ref{defZ}) from (\ref{defz}). Note also
that the magnetization vector of spin $i$ has components $m_i^\alpha =
\delta_{\alpha,1} m_i$.

The cavity method consists in adding a new spin ${\bf S}_0$ to
a $N$-site system $\{{\bf S}_1\ldots {\bf S}_N\}$. The new partition
function is calculated assuming that the statistical correlations in
the $N$-site system can be neglected. More rigorously we make use of
the {\em clustering theorem}, tacitly assuming that the system has
just one pure state. We can thus encode the effect of the whole system
onto each spin $i$ as an effective field $h_i$. The $N$+1-site
partition function can be written as:
\begin{equation}
\label{vectorspin}
Z_{N+1} = \lim_{p \rightarrow 0} \int \left[ \prod_{i=1}^N S_i^1
d\mu({\bf S}_i)\right] \exp \left( \sum_{i=1}^N h_i S_i + \sum_{i =
1}^N T_{0i} {\bf S}_0.{\bf S}_i \right)\,,
\end{equation}
and can be easily computed thanks to (\ref{loi}), giving:
\begin{equation}
\label{eq_tap}
m_0 = \Big( \sum _{i=1}^N T_{0i}m_i \Big)^{-1}\, ,
\end{equation}
where we have used the fact that $m_i$, the magnetization of site $i$
before the addition of site $0$ is $1/h_i$. At the end of the day we
are not interested in the spin variables but in the solution of the
matching problem. The thermal average of the occupation number of the
edge $0$--$i$ is simply related to the magnetizations by:
\begin{equation}
\label{occup}
\la n_{0i}\ra=m_0 T_{0i} m_i^c.
\end{equation}
Since we will be interested eventually in the ground state properties
of this model we have to take the $\beta\rightarrow\infty$ limit of
our equations. Following \cite{RatPar} it is useful to set 
\begin{equation}
\label{mtransf}
m_i = e^{\beta \phi_i} \,\,\, , \,\,\, \mathrm{for} \,\,\, i \in \{1,
\ldots , N\}\,.
\end{equation}
The zero temperature limit of (\ref{eq_tap}) and (\ref{occup}), thanks
to (\ref{mtransf}), reduces to the following zero-temperature cavity
relation:
\begin{eqnarray}
\label{eq_cav_mat}
\phi_0 &=& \min_{i=1,\ldots,N} N l_{0i} -\phi_i\,\,,  \\
n_{0i} &=& \delta_{i,i^*}\,\,,
\end{eqnarray}
where $i^*$ is the index attaining the minimum $Nl_{0i} - \phi_i$.

\subsection{The $\varepsilon$-coupling method}

The idea of the $\varepsilon$-coupling method in the context of the
matching problem is the following: given the set $\ell$ of $l_{ij}$
distances (called 0-distances), one first finds the minimum matching
(called 0-ground state), which is characterized by some occupation
numbers $n_{ij}$. Then one perturbs the lengths of the edges of the
graph by adding a quantity $\varepsilon$ to the edges present in the
0-ground state. Formally the edge-lengths become the following
$\varepsilon$-distances:

\begin{equation}
l_{ij}^\varepsilon=l_{ij}+\varepsilon n_{ij}.
\end{equation}
One solves the matching problem with these $\varepsilon$-distances and
obtains a solution we will name $\varepsilon$-ground state, which is
expected to be different from the 0-ground state. The larger
$\varepsilon$, the stronger the 0-ground state is penalized. The
$\varepsilon$-ground state is characterized by the occupation numbers
$n_{ij}^\varepsilon$. Two quantities are of interest: first the
difference of length (energy) between the $\varepsilon$-ground state
and the 0-ground state computed with the 0-distances:
\begin{equation}
\label{eq_Delta_L}
\Delta L_{\ell} = \sum_{i<j} (n_{ij}^\varepsilon-n_{ij})
l_{ij}.
\end{equation}
Second the distance $d$ between the $\varepsilon$-ground state and the
0-ground state:
\begin{equation}
d_{\ell}=1-q_{\ell},
\end{equation}
where the overlap $q_{\ell}$ is equal to the proportion of edges in
common:
\begin{equation}
q_{\ell}=\frac2N\sum_{i<j} n_{ij}^\varepsilon n_{ij}.
\end{equation}
More precisely one would like to compute the average with respect to
the coupling constants of these quantities, so let us define $\Delta L
= \overline{\Delta L_{\ell}}$ and $q = \overline {q_{\ell}}$. An
analytic approach is possible in the framework of the cavity
method. We have two spin systems, the 0-system and the
$\varepsilon$-system, standing on the same graph, and which are
coupled. More precisely the Hamiltonian of the $\varepsilon$-system is
conditioned by the ground state of the 0-system.

While the 0-system certainly obeys the cavity equations
(\ref{eq_cav_mat}), how to deal with the $\varepsilon$-system is more
problematic. A {\em naive} approach would lead us to:
\begin{equation}
\tau_0=\min_{i=1,\ldots N}\gamma_i,
\end{equation}
where 
\begin{eqnarray}
&& \gamma_i=l_{0i} - \tau_i \ \ \ {\rm if } \ \ i \ne i^*, \\ &&
\gamma_i=l_{0i}+ \epsilon -\tau_i \ \ \ {\rm if } \ \ i = i^*.
\end{eqnarray}
%\begin{figure}[htb]
%\begin{center}
%\includegraphics[width=10cm,angle=-90]{d_versus_eps.ps}
%\caption{illustration}
%\label{tralala}
%\end{center}
%\end{figure}
Nevertheless these equations are wrong! They would be true if the
interactions between the old spins in the $\varepsilon$-system of
$N+1$ spins were the same as in the $\varepsilon$-system of $N$ spins,
and this is not exactly the case: when adding the new spin to the
0-system, it gets matched to one of the old spins, whose previous
match becomes unmatched. To circumvent this difficulty, one should
distinguish between matched and unmatched spins in the 0-system. The
variable $\tau_i$ of the $\varepsilon$-system will be called $f_i$ if
the vertex $i$ is matched in the 0-system, and $v_i$ if it is not. The
correct equations are thus:
\begin{eqnarray}
\label{eqepssys}
&& v_0=\min_{i=1,\ldots N} (l_{0i}-f_i),\\
\label{second_case} && f_0=\min_{i=1,\ldots N} \gamma_i,
\end{eqnarray}
where 
\begin{eqnarray}
&& \gamma_i=l_{0i} - f_i \ \ \ {\rm if } \ \ i \ne i^{*}, \\
&& \gamma_i=l_{0i}+ \epsilon -v_i \ \ \ {\rm if } \ \ i = i^{*}.
\end{eqnarray}
The new spin in the $\varepsilon$-system gets matched to the spin
$i^{**}$, the index which minimizes Eq.(\ref{second_case}). The
contribution to the overlap $q$ is $\delta _{i^{*},i^{**}}$. The
contribution to $\Delta L$ is $l_{0i^{**}}-l_{0i^{*}}$.

\begin{figure}[htb]
\includegraphics[angle=0,width=0.95\columnwidth]{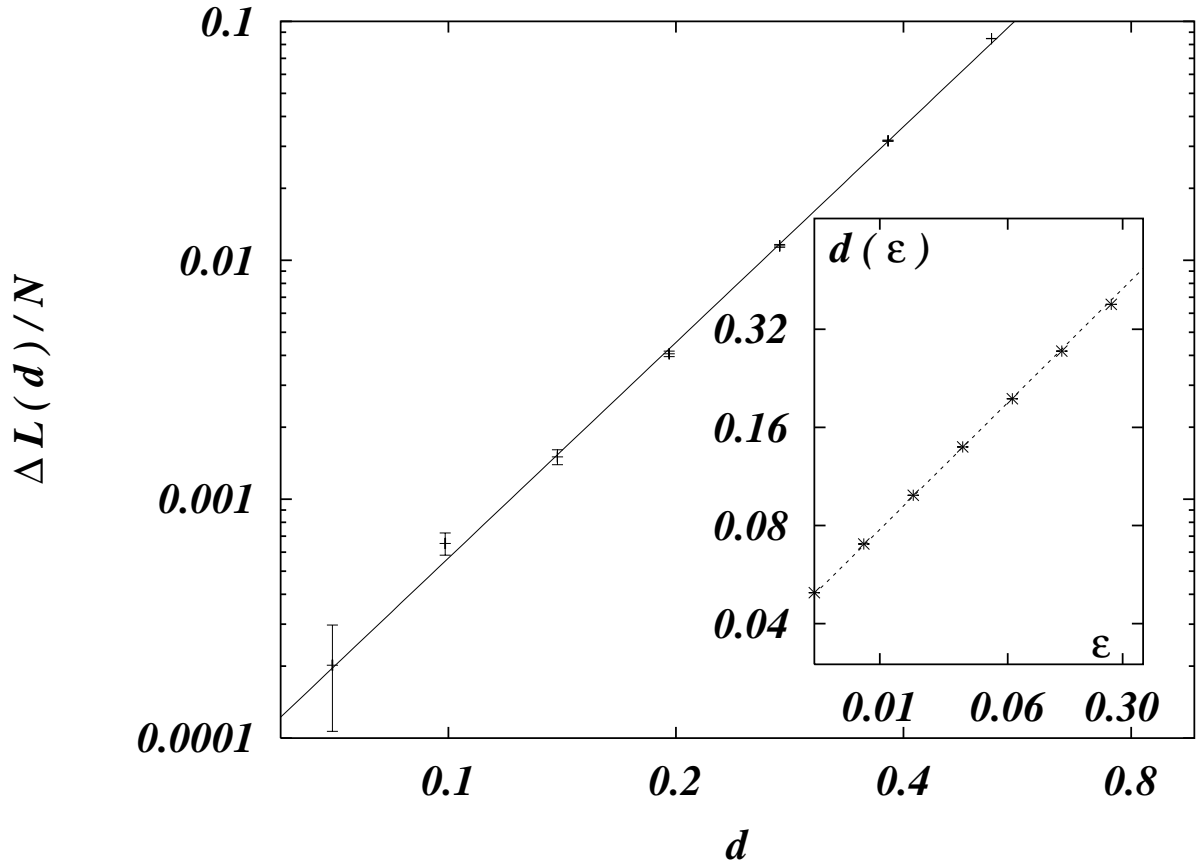}
\caption{Main panel: Average length $\Delta L/N$ {\sl vs.} distance
$d$, continuous line is the best fit of the form $const\times
d^3$. Inset: Average distance $d$ {\sl vs.} coupling parameter
$\epsilon$, dotted line is the best fit of the form $const\times
\sqrt{\epsilon}$.}
\label{fig:fits}
\end{figure}

When averaging over the disorder, in the thermodynamic limit
the quantities $(\varphi_0,v_0,f_0)$ and $(\varphi_i,v_i,f_i)$ are
i.i.d. random variables (beware that $\varphi$, $v$ and $f$ on the
same site are correlated). The above equations (\ref{eq_cav_mat}) and
(\ref{eqepssys}) define a stochastic flow whose fixed point is the
limit distribution of $(\varphi,v,f)$. We use a population algorithm
similar to the one discussed in \cite{MP01,MP02} to solve the
equations. In order to save computing time, we use a fluctuating
connectivity approximation of the matching problem: we keep the only
edges whose lengths are smaller than a given cutoff $z$, so that the
connectivity of a vertex is a Poisson r.v. of mean $z$. We store a
large population of $\cN$ triplets $(\varphi_i,v_i,f_i),\,i=1, \ldots
\cN$, which we initialize randomly and update iteratively: at each
step an integer $k$ is extracted following the Poisson distribution of
mean $z$, $k$ random elements of the population are chosen and we
compute $(\varphi_0,v_0,f_0)$ following the scheme defined by
Eqs.~(\ref{eq_cav_mat}), (\ref{eqepssys}) and (\ref{second_case})
limiting the search for the min to the $k$ extracted triplets; the
resulting $(\varphi_0,v_0,f_0)$ overwrites an element of the
population chosen at random. Once stochastic convergence of the
population is achieved, we keep on iterating and compute {\sl en
passant} the contributions to $\Delta L$ and $q$. Their flat averages
over many steps provide $\overline{\Delta L}$ and $\overline{d}$.
\begin{figure}[htb]
\includegraphics[angle=0,width=0.95\columnwidth]{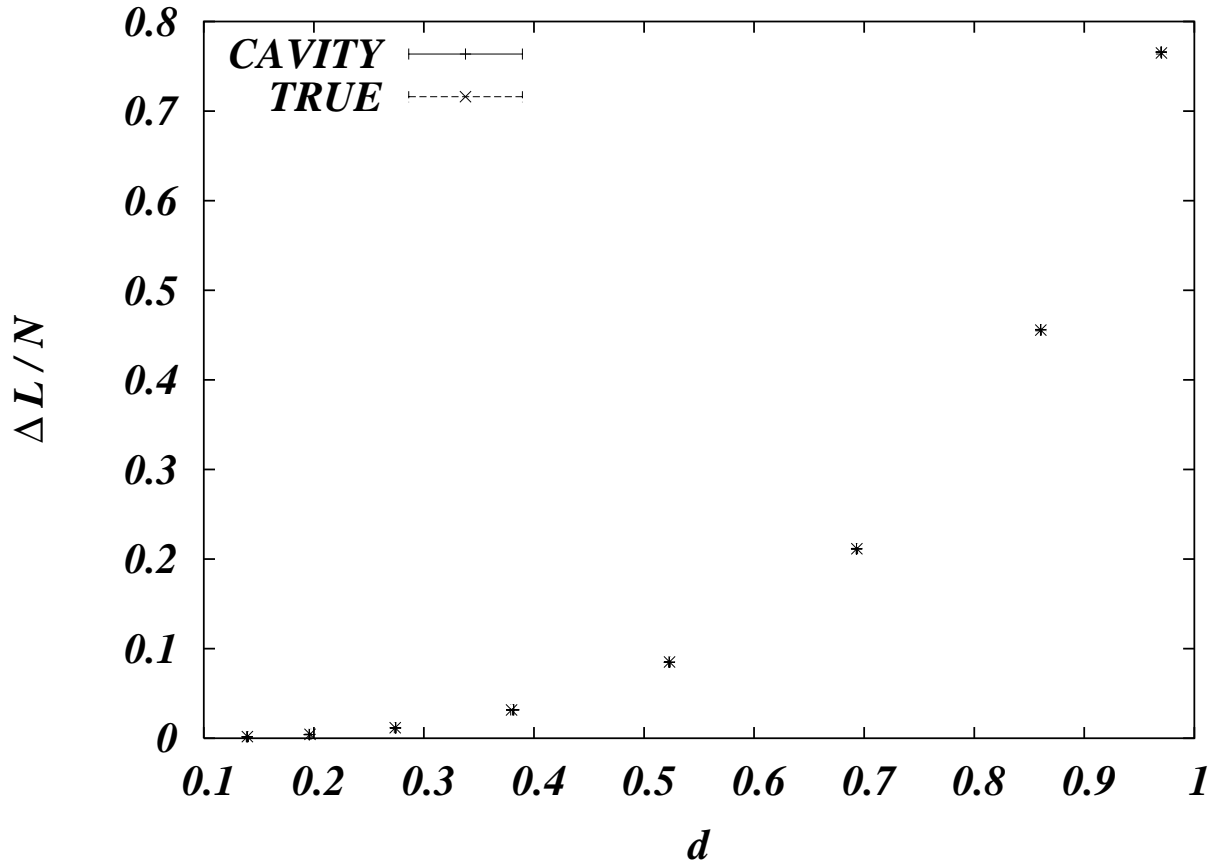}
\caption{$\Delta L / N$ {\sl vs.} $d$. The points obtained by the
cavity population algorithm with $z=30$, $\cN=200000$, and $10^8$
iterations (CAVITY), coincide with the measures obtained by averaging
over 4000 samples of total number of points $N=400$ (TRUE).}
\label{De_versus_d.ps}
\end{figure}
The output of this algorithm is presented in
Figs.~(\ref{De_versus_d.ps}) and (\ref{fig:fits}). Beside we have
calculated the ground state of 4000 matching instances using {\tt
BLOSSOM 4} software \cite{Algo_match} and we have tested the results
against the cavity approach. In Fig.~(\ref{De_versus_d.ps}) we display
$\Delta L/ N$ {\em vs.} $d$: the cavity approach is in perfect
agreement with the direct calculation. In Fig.~(\ref{fig:fits}) we
display the $\Delta L /N$ {\em vs.} $d$ curve (main panel) and the
best one-parameter fit of the form $const\times d^3$, while in the
inset we display the $d$ {\em vs.} $\epsilon$ curve together with the
best one-parameter fit of the form $const\times \sqrt{\epsilon}$: in
both cases the reduced $\chi^2$ is of order 1. Note that a simple
scaling argument shows that the two exponents are not independent: let
us assume that for $\epsilon$ small enough $d \sim \epsilon^\beta$,
then from Eq.~(\ref{eq_Delta_L}) $\Delta L/N \sim \epsilon
^{\beta+1}\sim d^{(\beta+1)/\beta}$. The scaling exponent $\alpha$
introduced by Aldous and Percus \cite{aldous_percus} is easily
recovered setting $\alpha = (\beta+1)/\beta$. We find that the random
matching features a scaling exponent $\beta =1/2$ ($\alpha = 3$), in
agreement with the results found in \cite{aldous_percus}. A completely
analytic study of the coupled system of Eqs.~(\ref{eqepssys}) and
(\ref{second_case}) might be done, but we did not undertake it.

An important remark is in order: in \cite{aldous} Aldous introduces
and proves the following {\em asymptotic essential uniqueness}
property:

\medskip
\noindent\begin{em}Let $M$ be the generic element of
a family of matchings depending both on $N$ and on the realization of
the $l_{ij}$; we call $q_{M,min}$ the overlap between $M$ and the
minimum matching, and $d_{M,min}=1-q_{M,min}$ their distance.

For each $0<\delta<1$ there exists $\varepsilon(\delta)>0$ such that:
if $\,\forall N,\,\,d_{M,min}\geq \delta$ then
\begin{equation}
\label{uniqueness}
\lim \,\inf_N \frac{L_M}{N} \geq
\frac{\pi^2}{6}+\varepsilon(\delta).
\end{equation}
\end{em}

In physical terms: in the thermodynamic limit, a matching
(configuration) which differs from the ground state by a non zero
proportion of edges has an energy density strictly greater than the
one of the ground state. The other way round: a state with the same
intensive energy density as the ground state can be obtained only by
changing a non extensive number of edges in the ground state.  This is
the proof that there is no RSB, at least at zero temperature. The plot
in Fig.(\ref{De_versus_d.ps}) can be seen as an illustration of this
theorem.

To conclude, note that our perturbation of the energy is $O(N)$ so the
information we get on the energy landscape is limited. In particular
we do not explore the lowest lying excited configurations which have
been numerically shown to have an energy $\Delta L\sim 1/\sqrt{N}$ and
$d\sim 1/\sqrt{N}$ \cite{HM98}. This is a limitation of our analytic
approach: it is purely thermodynamic so that we cannot have a hint at
finite size effects.

%% One could think of an alternative way of reducing the number of edges,
%% namely keeping only the $K$ edges with the smallest lengths incident
%% on each vertex. When $K \rightarrow +\infty$ this should give the same
%% result, but it is less convenient for algorithmic purposes, because it
%% is not easy to implement the probability law of the lengths of the $K$
%% smallest edges: either one picks the $K$ smallest among the total
%% $N-1$ edges, but this is time consuming, or one substitutes this law
%% by its limit when $K\rightarrow +\infty$, extracting independently $K$
%% numbers with the flat distribution on $[0,N]$. But corrections are in
%% $O(1/K^\nu)$.

\section{The Gaussian spin glass on the Bethe lattice}
\label{sec:gaussian_SG}
In close analogy with the previous section, we derive cavity equations
for the $\varepsilon$-coupling method applied to the Gaussian spin
glass on the Bethe lattice of connectivity $k+1$, in the presence of
an external field. We keep to the level of the RS approximation. We
will see that our results provide a self-consistency check of this
hypothesis which enables to trace out when it is valid or not.

Throughout this section we make thorough use of the notations introduced in our previous article \cite{previous}. We refer to sections II and III therein for details. 

\subsection{The cavity equations for a single system at zero temperature}

\label{The cavity equations for a single system at zero temperature}

\begin{figure}
\includegraphics[angle=0,width= 0.95\textwidth]{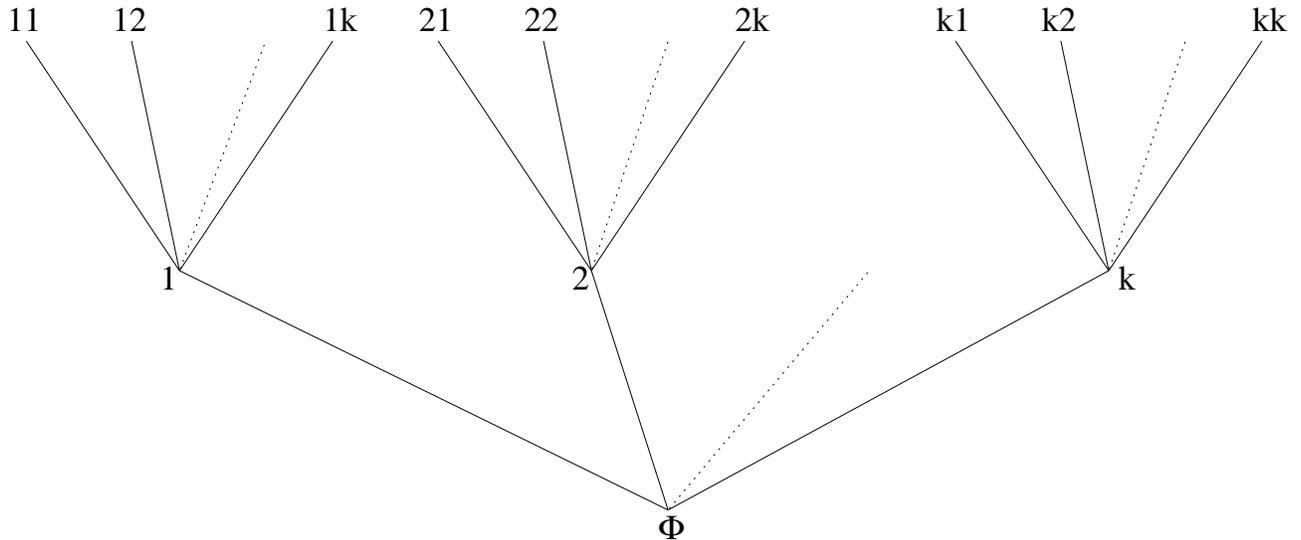}
\caption{A branch of a Cayley tree of connectivity $k$. %Note that each
% vertex, including the root, has $k$ sons.
}
\label{tree}
\end{figure}

First, following M\'ezard and Parisi \cite{MP02} we work out the
cavity method directly at zero temperature for a single system of $N$
spins $\s$ interacting through coupling constants $J$ -- their set being denoted $\mathcal J$ -- which are i.i.d. random
variables with a Gaussian distribution of mean 0 and variance 1, in
the presence of an external field $h_{ext}$.

Let us consider the merging process of $k$ branches rooted at spins
$i=1,\ldots k$ onto a new spin $\Phi$ illustrated in
Fig.~(\ref{tree}). We look at how the energy of the ground state
evolves under this process. Before the merging, on a given branch, we
let the spin at the root undetermined. Thus the ground state energy of
the branch rooted at spin $i$ is conditioned by the value of the spin
$i$ and can be written:
\begin{equation}
\label{a branch before}
E(\s_i)=A_i-h_i \s_i,
\end{equation}
where $A_i$ is a constant, and $h_i$ is an effective field (beware
that it is not the local field). Note that $h_i$ contains the effect
of the external field $h_{ext}$. As a consequence, the energy of the
system of the $k$ branches before the merging is conditioned on the
values of the spins $1,\ldots k$, and reads
\begin{equation}
E(\s_1,\ldots \s_k)=A_1+\ldots+ A_k-h_1 \s_1-\ldots-h_k \s_k.
\end{equation}
The system after the merging has an energy
\begin{equation}
E'(\s_1,\ldots \s_k,\s_\Phi)\!=\!A_1\!+\!\ldots
\!+\!A_k\!-\!(h_1\!+\!J_{\Phi,1}\s_\Phi)\s_1\!-\!\ldots\!-\!(h_k\!+\!J_{\Phi,k}\s_\Phi)\s_k\!-\!h_{ext}\s_\Phi,
\end{equation}
which, in order to be that of the ground state, is to be minimized with
respect to $\s_1,\ldots \s_k$ at fixed $\s_\Phi$. This is realized by
independently choosing the sign of each $\s_i$ such that
$(h_i+J_{\Phi,i}\s_\Phi)\s_i=|h_i+J_{\Phi,i}\s_\Phi|$. As we can write
\begin{equation}
|h_i+J_{\Phi,i}\s_\Phi|=\omega(h_i,J_{\Phi,i})+\lambda(h_i,J_{\Phi,i})\s_\Phi,
\end{equation}
where
\begin{eqnarray}
&& \omega(h,J)=\frac{|h+J|+|h-J|}{2},
\\\nonumber && \lambda(h,J)=\frac{|h+J|-|h-J|}{2}.
\end{eqnarray}
we get the appealing following form:
\begin{equation}
E'(\s_\Phi)=A_1+\ldots+A_k-\sum_{i=1}^k
\omega(h_i,J_{\Phi,i})-\left(\sum_{i=1}^k
\lambda(h_i,J_{\Phi,i})+h_{ext}\right)\s_\Phi\,\,.
\end{equation}
By comparison with Eq.(\ref{a branch before}) this leads to the
recursion relation:
\begin{equation}
h_\Phi=\sum_{i=1}^k \lambda(h_i,J_{\Phi,i})+h_{ext}.
\end{equation}

\subsection{The cavity equations for two coupled systems}
\label{The cavity equations for two coupled systems}
\begin{figure}[htb]
\includegraphics[angle=0,width= 0.95\textwidth]{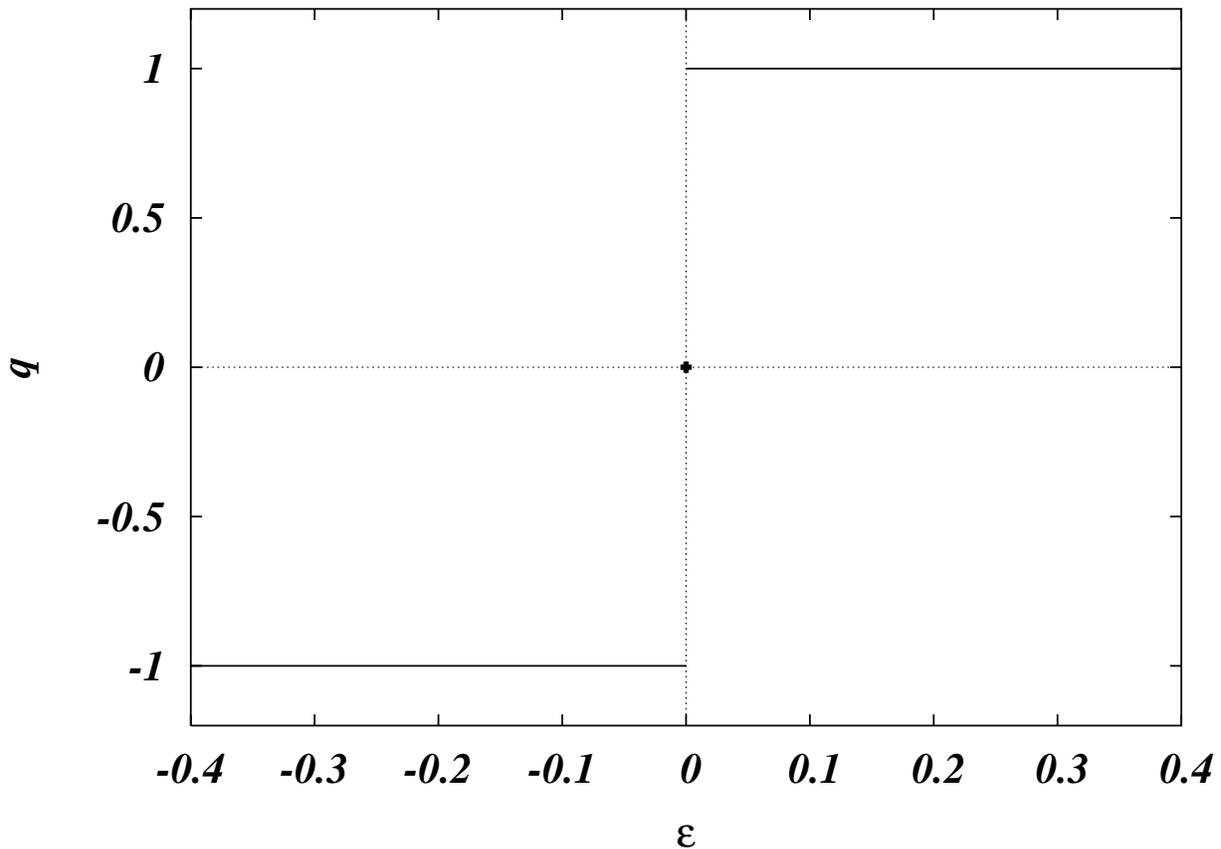}
\caption{$q$ {\sl vs.} $\varepsilon$ for $k=2$, in the case
$h_{ext}=0$.}
\label{q_versus_eps_h0.ps}
\end{figure}

\begin{figure}[htb]
\includegraphics[angle=0,width= 0.95\textwidth]{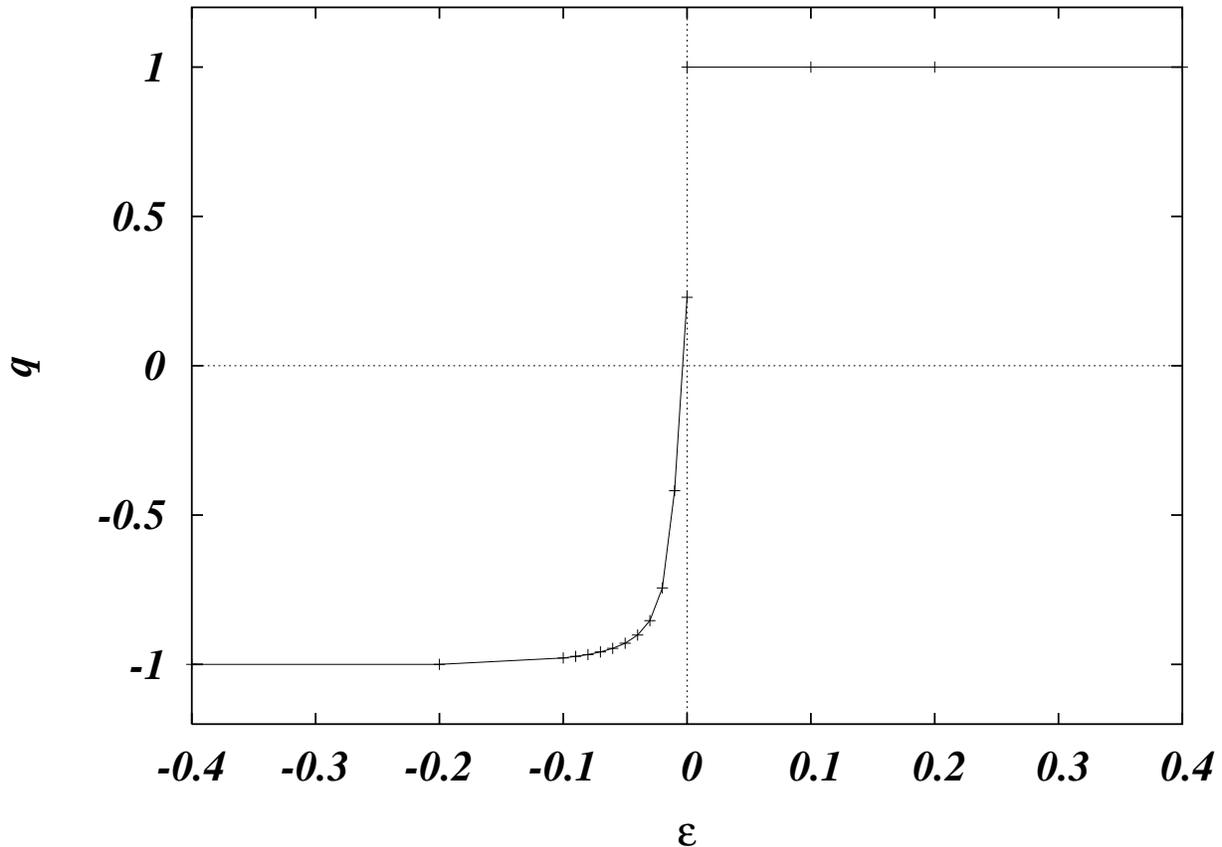}
\caption{$q$ {\sl vs.} $\varepsilon$ for $k=2$, in the case
$h_{ext}=0.1$, obtained with $\cN=2000$ and $100000\times \cN$
iterations.}
\label{q_versus_eps.ps}
\end{figure}

\begin{figure}[htb]
\includegraphics[angle=0,width= 0.95\textwidth]{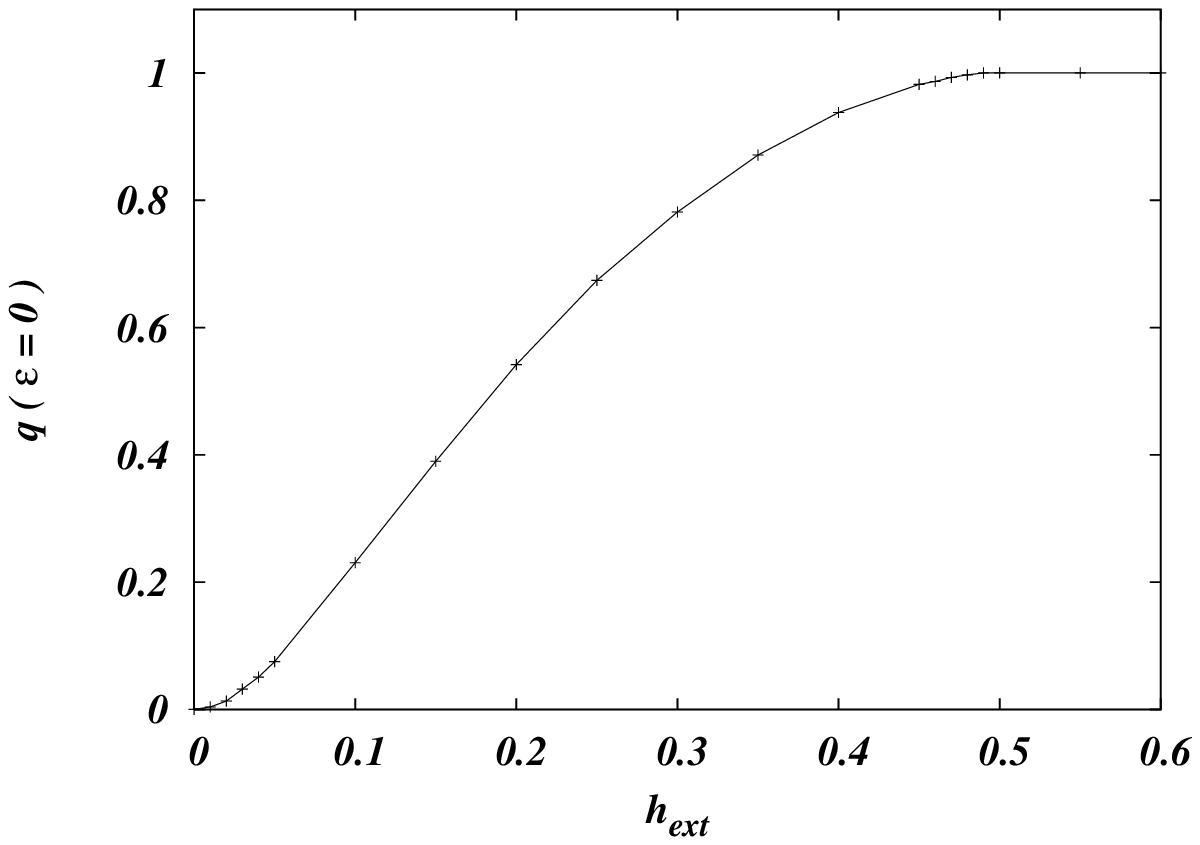}
\caption{$q$ {\sl vs.} $h_{ext}$ for $k=2$,
$\varepsilon=0$, obtained with $\cN=2000$ and $100000\times \cN$
iterations.}
\label{q_versus_h.eps}
\end{figure}

Now we imagine there are two spin systems sitting on the $N$ vertices
$A$ of the {\sl same} Bethe lattice:
\begin{itemize}
\item the $\s$ spins obey the same Hamiltonian as in the previous
section \ref{The cavity equations for a single system at zero temperature}:
\begin{equation}
\label{eq_ham_s}
H[\sigma] = - \sum _{\la A,B \ra} J_{A,B} \sigma_A \sigma_B-h_{ext}
\sum_A \s_A.
\end{equation}
Let us call $\s^*=(\s_A^*)$ the ground state of this Hamiltonian.
\item the $\tau$ spins obey the following perturbed Hamiltonian,
conditioned on the $\s_A^*$:
\begin{equation}
\label{eq_ham_t}
H[\tau|\sigma^*] = - \sum _{A,B} J_{A,B} \tau_A \tau_B -h_{ext} \sum_A
\tau_A-\epsilon \sum_{A} \sigma_A^*\tau_A.
\end{equation}
The choice $\varepsilon>0$ corresponds to an attractive interaction
for the $\tau$ spin variables to the configuration $\s^*$, the choice
$\varepsilon<0$ to a repulsion. The ground state of the Hamiltonian
(\ref{eq_ham_t}) is named $\tau^*=(\tau_A^*)$.
\end{itemize}

We are interested in the overlap
\begin{equation}
q \equiv \overline {q_{\mathcal J}}=\frac{1}{N}\sum_i \overline{ \s_i^*
  \tau_i^*} \,\,.
\end{equation}
where with the symbol $\overline{O_{\mathcal J}}$ we mean the
average of a generic ${\mathcal J}$-dependent observable $O_{\mathcal
J}$ over the different sample realizations.  The correlations between
the $\s$ spins and the $\tau$ spins can be accounted for by stating
that the value of the effective field acting onto spin $\tau_\Phi$
depends on the value of spin $\s_\Phi$, where $\Phi$ is the
root of an isolated branch. So we have to store three quantities
related to the root $\Phi$ of a branch:
\begin{itemize}
\item[-] $h_\Phi$, the effective field acting onto spin $\s_\Phi$;
\item[-] $h_\Phi^+$, the effective field acting onto spin $\tau_\Phi$
under the condition that spin $\s_\Phi$ has the value +1;
\item[-] $h_\Phi^-$, the effective field acting onto spin $\tau_\Phi$
under the condition that spin $\s_\Phi$ has the value -1.
\end{itemize}

\subsubsection{Iteration}

When performing the merging of $k$ branches onto a new site, the
equation for the system of the $\s$ spins is exactly the same as in
the previous section \ref{The cavity equations for a single system at
zero temperature}:
\begin{equation}
h_\Phi=\sum_{i=1}^k \lambda(h_i,J_{\Phi,i})+h_{ext}.
\end{equation}
As far as the $\tau$ spins are concerned, one must compute $h_\Phi^+$
as well as $h_\Phi^-$. So we assume in turn that $\s_\Phi=+1$ and
$\s_\Phi=-1$. This determines the values of the $\s_i^*$:
\begin{equation}
\label{sigma_i}
\sigma_i^*=\textrm{sign}(h_i+J_{\Phi,i}\sigma_\Phi^*),
\end{equation}
and the effective field acting onto spin $\tau_\Phi$ reads
\begin{equation}
h_\Phi^{\sigma_0^*}=\sum_{i=1}^k
\lambda(h_i^{\sigma_i^*},J_i)+\varepsilon \sigma_\Phi^*+h_{ext}.
\end{equation}
\subsubsection{Measures}
To measure the overlap, one uses the merging procedure of $(k+1)$
branches onto a new vertex $\Psi$: the effective field acting onto
spin $\s_\Psi$ is:
\begin{equation}
H_\Psi=\sum_{i=1}^{k+1} \lambda(h_i,J_{\Psi,i})+h_{ext}\,.
\end{equation}
By contrast with the iteration, spin $\s_\Psi$ has a determined value,
the one which minimizes the energy:
$\sigma_\Psi^*=$sign$(H_\Psi)$. This also determines the values of the
$\sigma_i^*$ according to Eq.~(\ref{sigma_i}). Eventually one can
compute the effective field acting onto spin $\tau_\Psi$:
\begin{equation}
H_\Psi=\sum_{i=1}^{k+1} \lambda(h_i^{\sigma_i^*},J_i)+\varepsilon
\sigma_0^*+h_{ext},
\end{equation}
whose sign gives $\tau_\Psi^*$. The contribution to the overlap is
$\sigma_\Psi^* \tau_\Psi^*$.

\subsection{The results}

Unless otherwise stated the following results are for $k=2$. Once
again they have been obtained by a population algorithm. Here the
population is made of $\cN$ triplets $(h,h^+,h^-)$.

In the case $\varepsilon >0$ we expect $q$ to be equal to 1,
regardless of the value of $h_{ext}$: the ground state of the
perturbed Hamiltonian (\ref{eq_ham_t}) should be the same as the
Hamiltonian (\ref{eq_ham_s}). We will see that so it is. Thus the
interesting regime is $\varepsilon \leq 0$.

What we obtain in the case $h_{ext}=0$ is plotted in
Fig.(\ref{q_versus_eps_h0.ps}). The fact that $q=-1$ for every
$\varepsilon <0$ is due to the symmetry of the original Hamiltonian
(\ref{eq_ham_s}) under reversal of all the spins: the ground state of
the perturbed Hamiltonian is $\tau_A^*=\s_A^*$. More interesting is
the fact that at $\varepsilon=0$, $q=0$: assuming RS one would expect
$q=1$. This is a sign that the RS ansatz is not self consistent, and
is to be dismissed.

Things may be more convincing if one turns the external field on. The
case $h_{ext}=0.1$ is plotted on Fig.(\ref{q_versus_eps.ps}). As the
symmetry of the original Hamiltonian is lifted, the plot of $q$ as a
function of $\varepsilon<0$ is no longer a constant equal to -1. It
does however tend to -1 when $\varepsilon \rightarrow -\infty$ because
in this limit the attracting term between the $\s$ system and the
$\tau$ system dominates the perturbed Hamiltonian
(\ref{eq_ham_t}). When $\varepsilon \rightarrow 0^-$ $q$ goes
continuously to a value which is no longer 0, but is still not
1. Again the RS ansatz is in trouble.

We do expect however the RS ansatz to be valid for a sufficiently high
external magnetic field. Fixing $\varepsilon=0$, we let $h_{ext}$
increase. See Fig.(\ref{q_versus_h.eps}): it appears that $q$ is an
increasing function of $h_{ext}$ and saturates to 1 at $h_{ext}^c \sim
0.48$. It is the sign that the RS ansatz becomes self-consistent above
$h_{ext}^c$.

For $k=5$ we find $h_{ext}^c \sim 1.86$ which is different from the
value $h_{ext}^c \sim 2.1$ found in \cite{HM-magneticfield} by the
analysis of numerical simulations. We believe our result is exact, and
the discrepancy can be explained by the fact that their result relies
on finite size scaling arguments with a relatively poor precision.

What we got here is actually the point at $T=0$ of the
de~Almeida-Thouless (AT) line \cite{at} for the Gaussian spin glass on
the Bethe lattice. We had the idea we could generalize our approach to
a non zero temperature to determine the whole line.

\subsection{Searching for the AT line}

It is not difficult to generalize the argument of section 
\ref{The cavity equations for two coupled systems} (in the only case
$\varepsilon=0$) to a non zero temperature. We consider two non
interacting systems, the $\s$ spins and the $\tau$ spins, standing on
the {\sl same} Bethe lattice and obeying the {\sl same} Hamiltonian
(\ref{eq_ham_s}). Solving this double system at the level of RS is
easy: we use the population algorithm of section \ref{The cavity
equations for a single system at zero temperature} adapted to follow
simultaneously two populations. A crucial point is that whenever one
randomly extracts sites or coupling constants, they are the {\sl same}
for the $\s$ population and the $\tau$ population. This procedure
enables to measure the average overlap between the two systems
\begin{equation}
q\equiv \overline{q_{\mathcal J}}= \overline{\la \s_A \ra \la \tau_A \ra}.
\end{equation}
The criterion for the RS ansatz to be self-consistent is
\begin{equation}
\label{criterion for the RS ansatz}
q = \overline{m_A^2},
\end{equation}
where $m_A$ is the local magnetization measured either in the
$\s$-system or the $\tau$-system (they are obviously equal).

\begin{figure}[htb]
\begin{center}
\includegraphics[angle=0,width= 0.95\textwidth]{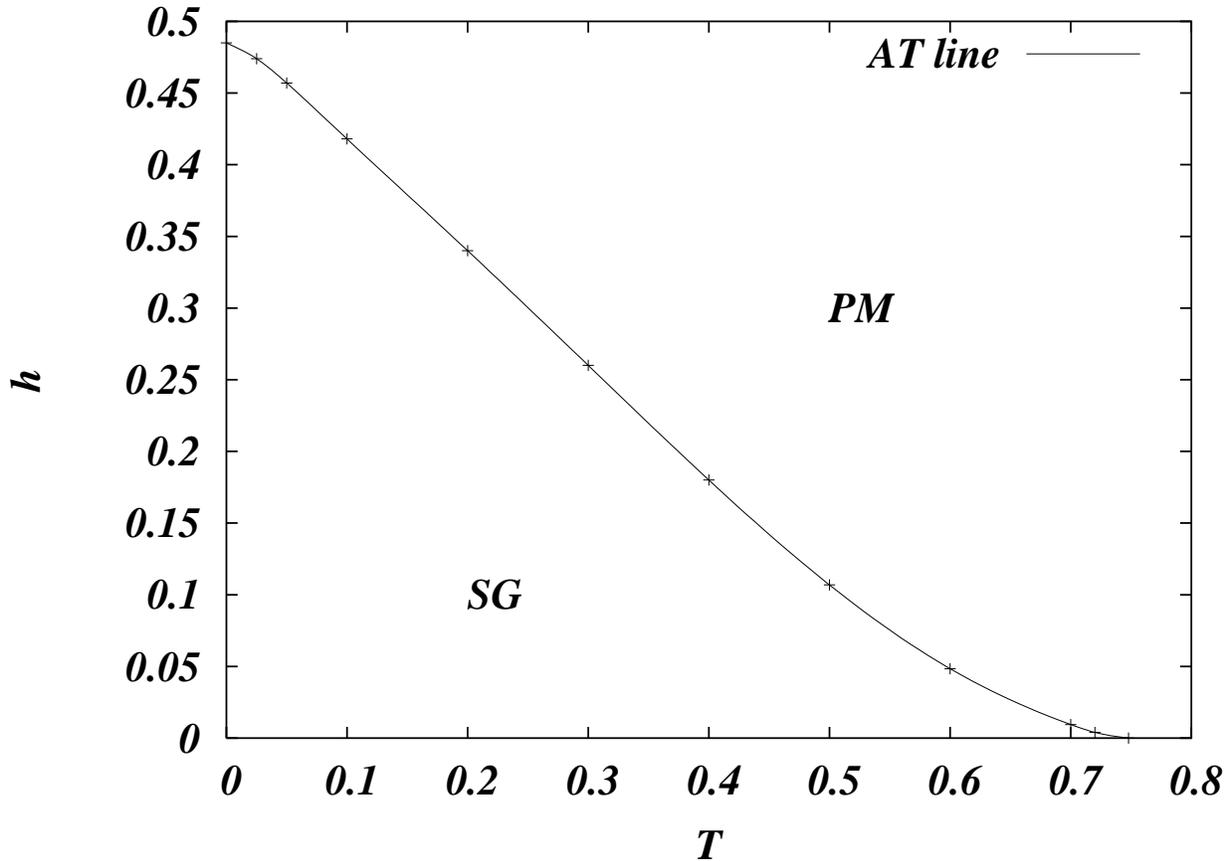}
\caption{The de~Almeida-Thouless line ($h$ {\sl vs.} $T$) on the
Bethe lattice in the case $k=2$, separating Spin-Glass (SG) phase from
the Paramagnetic (PM) phase. Data are obtained with $\cN=10000$ and
$10000\times \cN$ iterations. We estimate error bars to be hardly
visible on this scale. Continuous line is just a guide to the eye.}
\label{ATline.ps}
\end{center}
\end{figure}

Given a value of $T$, we run the algorithm for increasing values of
$h_{ext}$, so as to determine its value $h_{ext}^c(T)$ beyond which
condition (\ref{criterion for the RS ansatz}) holds. The plot
$h_{ext}^c(T)$ is the AT line. See Fig.(\ref{ATline.ps}). To our
knowledge it is the first time the AT line has been obtained for a
spin glass on the Bethe lattice. Two predictions made in
\cite{thouless} can be checked. First, the critical temperature $T_c$,
such that $h_{ext}^c(T_c)=0$, is the solution of the equation
$k\,\overline{\tanh^2(J/T_c)}=1$: in the case $k=2$ this gives
$T_c=0.748$. Second close to $T_c$, $h_{ext}^c(T)$ should behave like
$(T_c-T)^{3/2}$, which a numerical fit of our data confirms.

\section{Conclusions and perspectives}
\label{sec:concl}
In this paper we presented the derivation and implementation of the
$\varepsilon$-coupling method in the framework of the cavity
approach. This technique allows us to explore the way the energy of a
configuration varies as a function of its overlap with the ground state
and more generally to address the problem of the organization of the
lowest energy configurations. The lesson we get applying this method
to the case of the simple random matching problem is very clear: the
space of the lowest energy configurations is organized such that their
energy difference with respect to their distance from the ground state
scales as $\Delta E / N \propto d^3$. A situation like this, or in
general whenever $\Delta E / N \propto d^\alpha$ with $\alpha > 0$,
is related to the property of replica symmetry of the system, which
implies Aldous {\em asymptotic essential uniqueness} property
\cite{aldous}.

A similar study presented in \cite{aldous_percus} suggest that this
property is also shared by the Minimum Spanning Tree problem, the
Minimum Matching Problem in Euclidean dimension $d=1$, and the
Traveling Salesman Problem also in Euclidean dimension $d=1$ (all with
$\alpha = 2$). Minimum Matching Problem and Traveling Salesman Problem
in $d=2,3$ are instead characterized by $\alpha = 3$ as the mean field
matching problem we have studied.

A simple case with no {\em asymptotic essential uniqueness} property
is the spin glass on a fixed connectivity random graph studied in
Sec.~\ref{sec:gaussian_SG}.  Indeed our computation based on the RS
assumption yields that $\lim_{\epsilon\rightarrow 0^{-}} d \neq 0$
which is a physical nonsense since the model has a unique ground state
(the couplings are gaussian). This inconsistency tells us not only
that - as we already know after \cite{MP01,MP02} - the cavity
approximation must be improved in order to take into account the
presence of many states, but also gives us a practical tool to probe
the phase space for the onset of full RSB: the search for the AT line
in the case of the spin glass on a fixed connectivity random graph, is a
simple and instructive example.

A very interesting issue is the generalization of the
$\varepsilon$-coupling method to the case where the {\em asymptotic
essential uniqueness} does not hold. In the last year a compact and
efficient formalism has been developed to apply the cavity method to
SAT and Coloring Problems \cite{ksat, coloring}, at the level of 1RSB,
evidentiating a clustering transition.  This clustering transition
consists in the sudden appearance of an exponential number of
metastable states, which - intuitively - cause local search algorithm
to get stuck. We believe that it is possible to generalize the
$\varepsilon$-coupling method presented in this paper to the 1RSB
level, although at a higher computational cost. This could give
interesting results on the inner mechanism of the clustering of
states. Open problems like how large a single cluster could be and
what the inter-cluster mean distance is, could be addressed within
this formalism.

\begin{acknowledgments}
We are grateful to F. Ricci-Tersenghi for many interesting
discussions. AP acknowledges the financial support provided through
the European Community's Human Potential Programme under contract
HPRN-CT-2002-00307, DYGLAGEMEM.
\end{acknowledgments}

\end{document}